\documentclass[twocolumn,showpacs]{revtex4}
\usepackage{graphicx}
\usepackage{dcolumn}
\usepackage{bm}
\usepackage{amssymb}

\begin{document}

\preprint{}

\title{The Origin of CP Violation}

\author{J.C. Yoon}
\email{jcyoon@u.washington.edu} \affiliation{University of
Washington}

\date{\today}

\begin{abstract}
Quantum field theory has been established on symmetries, but their
fundamentality could be limited as practical calculations of
physical observations are not based on interacting Lagrangian. The
requirement of Lorentz invariance on vacuum expectation values is
contradicted by parity violation with massive fermions and Lorentz
violation of the approximation in the standard model is proved.
After alternative interpretation of quark mixing and problems of
CP violation are addressed, the composite properties of neural
meson will be suggested as origin of CP violation with possible
experimental tests. In conclusion, fundamental symmetry violation
is still inconclusive due to the limited theoretical assumptions
and physical observations.
\end{abstract}

\maketitle
\section{Introduction.}
Symmetry has been playing an important role in the establishment
of quantum field theory as a fundamental principle to understand
interactions of elementary particles. However, its fundamentality
has never been examined thoroughly even when symmetry violation
was suggested. There could be a fundamental theory that really
governs physics in nature, however, it would be also reasonable
and more rigorous to study physics only based on what we can
verify through physical observations without {\it a priori} of
theoretical fundamentality and ambiguous analogies. Here let us
limit its {\it fundamentality} by physical observations and
investigate the fundamental principle of symmetries. This
perspective will lead us to view discrete symmetries and their
violations with more care and open to other possible
interpretations in elementary particle physics.

\section{The fundamentality of Symmetries}
Fundamentality of a theory can be discussed from equations it is
founded on and evaluated on its connections to physical
observations. Interacting Lagrangian has been considered to be
fundamental describing field interactions, and its symmetry has
been studied to establish quantum field theory. However, practical
calculations of physical observables are not based on this
interacting Lagrangian but on the assumption of free fields and
their local interactions. This mathematical assumption is that
quantum states after interactions are closely related to their
initial ones, which ends up with the assumption of free fields in
infinite time due to the locality of
interactions\cite{LippmanSchwinger,GellMannGoldberger}. The decay
rates, for example, would be the same regardless of relatives
signs of interaction terms to free field equations. The
interactions in this approach should be considered as independent
physical events separated from free fields. This point becomes
clear with the consideration of more than one interactions. Though
they could be included in Lagrangian together, generally what they
practically mean is not that these interactions occur in the same
physical event but that they represents possible interactions as
separated physical events, which can be verified with decay rate
calculations; the lifetime of particle is independent of the
relative signs between them\cite{LeeYang57}. Since physical
observables are the same even if Lagrangian is not invariant under
symmetries, symmetries of interacting Lagrangian is limited to be
fundamental. If it is fundamental, parity violation of weak
interactions contradicts the unification of electroweak
interactions since it implies fundamental difference between these
two interactions, which was claimed to be resolved by the standard
model but it violates Lorentz invariance as explained later. Its
fundamentality is also limited by the arbitrariness of discrete
symmetries, since they are defined from one specific interactions
of electromagnetic interactions while we could have arbitrarily
defined symmetry with weak interactions and found some symmetries
of electromagnetic interactions violated.

Local gauge theories, based on interacting Lagrangian, are also
limited to be fundamental since the invariance of Lagrangian is
not required for the practical calculations of the observations
and there is no physical evidences for gauge transformations or
local gauge, which is at best an excessive degree of freedom. The
gauge invariance was introduced without any physical consequences
of the local gauge and gauge transformations, but it was accepted
without any doubt even when new fields and interactions were
introduced in the gauge transformations\cite{YangMills}. When the
assumption that Lagrangian is invariant under gauge transformation
is confronted with fermion mass problem, instead of investigating
its fundamentality more unfounded assumptions of spontaneous
symmetry breaking and Higgs boson were employed to resolve this
failure\cite{Higgs,Kibble}. This resolution, however, made its
physical evidences more elusive than before; even if Higgs boson
is found, this would not be enough to verify the fundamentality of
spontaneous symmetry breaking. Since the local gauge symmetry is
unnecessary in theory and unfeasible in experiments, fundamental
interactions should be investigated rather empirically than by
local gauge theories.

Therefore, physical observables of interactions obtained from the
practical calculations are irrelevant whether Lagrangian is
invariant under symmetries and thus symmetries of Lagrangian is
not fundamental to physical observations.

\section{CPT Symmetry}
When we observe particle and antiparticle symmetry through mass
and lifetime equality, this observations are made for particle and
antiparticle respectively, not involving any kind of physical
transformations from a particle to its antiparticle. Since
symmetry transformations are not physical ones and it is only an
assumption that they transform a field into its corresponding
field, its experimental verification is unfeasible. The effort to
identify particle and antiparticle symmetry as an symmetry
transformations is subjected to more fundamental concept of the
mass and lifetime equality that is grounded for determining
fundamental equations. The fundamental equations, for example,
Dirac equations for particle and antiparticle are determined from
the mass and lifetime equality and charge conjugation(C), the
relation between a field and its antifield, is determined
accordingly. However, the mass and lifetime equality should be
applied only to particles not to fields that include quantum
states besides the properties of particle, since mass and lifetime
of a particle are the same regardless of its quantum states such
as spin orientation and helicity if massive. Though particle and
antiparticle symmetry should be irrelevant of quantum states, C,
supposed to be this symmetry, is defined on fields, not on
particles in order to incorporate with time reversal(T) that
concerns with quantum states considering spin orientations under
the reversed flow of time, which is only a theoretical assumption.
Therefore, C, and thus CPT, is sufficient but not necessary for
the particle and antiparticle symmetry. Since the time reversed
spin orientation and the antilinearity of T, introduced to obtain
the desired commutation relations of boson\cite{Luders}, do not
provide any physical evidences, it is reasonable to consider T and
C arbitrarily defined to satisfy CPT theorem. The interpretation
that an antiparticle is the time reversed particle is induced from
the interactions of quantum electrodynamics, but this could be
contradicted by weak interactions with different fundamental
particles involved. This interpretations should be accepted as a
fundamental assumption that is limited to its mathematical
descriptions. Time reversal in particle physics is applied to
interactions by exchanging initial and final states of
interactions rather than to particles by physically reversing the
flow of time. Though we observe such a symmetry, it does not
necessarily mean T invariance since it would be symmetric as long
as the interactions are independent of time or instantaneous. Most
of T violation claimed in experiments could be interpreted as
physical effects from electric dipole moment of nuclei, which is a
property of composite particles from charge
distributions\cite{GingesFlambaum}.

Parity(P) has been considered to be different from Lorentz
invariance and helicity has been considered as a fundamental
property of particle since what it does to the helicity of
massless fermion, neutrino, could not be obtained by Lorentz
transformations and that of massless fermion is expected to be the
same under Lorentz
transformations\cite{LY57Neu,Barger,Buchmuller}. However, this
should not have been generalized to massive fermions since Lorentz
transformations can change their helicities as parity does and
thus Lorentz invariance implies parity of massive fermion is
conserved. The helicity of massive fermion is not a fundamental
property that defines particle since it is not conserved under
Lorentz transformations. It would be also excluded for neutrino if
massive
\cite{SuperKSmy,Hallin03,SuperKEguchi,GALLEX,GNO,SAGE,Cleveland}.
Though the change of helicity in neutrino would be extremely hard
to be observed due to its small mass, neutrino should be the same
particle whether it is left-handed or right-handed as long as
neutrino has nonzero mass. Parity should be considered as one kind
of Lorentz invariance especially when there is no massless fermion
and thus Lorentz invariance of vacuum expectation values is
contradicted by parity violation.

In particle physics, we apply the idea of parity to composite
particles such as mesons inadvertently. A meson, for example, is
supposed to have parity from its orbital angular momentum and an
additional sign from its fermion exchange. This additional sign
could be explained from quantum mechanics, but it is incorrect
since quantum mechanical system is required to have identical
particles to be either symmetric for bosons or antisymmetric for
fermions. Its Hamiltonian is not symmetric under the exchange of
two particles because fermions in a meson are different in masses
and charges and thus meson is not an eigenstate of parity. The
application of this fermion exchange seems to be valid because
either one or three fermions are obtained from fundamental
interactions of an initial fermion. The failure of this
approximation can be observed in such decays that involve quark
hadronizations contributing another sign as in $\tau-\theta$
puzzle. This is because meson is not a point particle but a
composite particle so that its decay could involve more than a
fundamental interactions.

In the standard model, weak interactions of massive fermions are
described as those of fermions with definite helicities in the
massless limit\cite{Glashow,Weinberg,Salam}, but this violates
Lorentz invariance while the fundamentality of helicity is
subjected to its approximation. Taking the massless limit
neglecting depressed helicity states in the boosted frame is
Lorentz violating, since the exact quantum states in the rest
frame cannot be achieved by Lorentz transformations of the
approximated ones back to the rest frame, for example,
\begin{eqnarray}
\sqrt{2E} \left( \begin{array}{c}  \xi \\ 0 \end{array} \right)
\nrightarrow \sqrt{m} \left( \begin{array}{c} \xi \\ \xi
\end{array} \right) \nonumber
\end{eqnarray}
where $\xi$ is an arbitrary two component spinor. No massive
fermion can be represented with only one definite helicity in the
boosted frame without Lorentz violation and the mixed states of
helicity should have been carefully considered in experimental
tests for parity violation\cite{Derman,E158}. Since Lorentz
invariant observables should be the same regardless of the frame
of reference, there is no good reason to perform their
calculations in the boosted frame approximately when exact
calculations are available in the rest frame and helicity and
chirality of massive particles are not well defined neither under
Lorentz transformations nor in the rest frame. This approximation
allows to view weak interactions as the interactions of fermions
with definite helicity, but they were originally interpreted as
interactions with a certain structure($1 \pm \gamma^5$), which is
well defined in the rest frame\cite{FeynmanGellMann,Sudarshan}.
Unlike the proper interpretation, this approximated one makes the
unification of electroweak interactions possible avoiding the
difference between these interactions in the fundamental symmetry
violation of parity, but this should not have been justified
because of its Lorentz violation.

The conclusion of parity violation is also premature due to the
neglect on the approximation of composite particles to point
particles and inaccurate concept of parity from the assumption of
massless neutrino. The beta decay of nuclei\cite{CSWu} was
suggested as an experimental proof for parity violation. What they
called pseudoscalar term is a product of $\it electron$ momentum
and spin of $\it nuclei$. When they found it asymmetric under
parity, they concluded that parity was violated\cite{LeeYang56}.
However, it is of question whether this interaction is fundamental
since nuclei is assumed to be a point particle ignoring its charge
distribution. The electric dipole moment of nuclei, claimed to be
induced by this symmetry violation, could also be interpreted
simply as a composite property of nuclei from its charge
distribution. The parity violation in the experiments on $\pi
\rightarrow e + \nu$ decays \cite{Garwin,Telegdi} would be
contradicted if neutrino is massive. Helicity of massive particle
is neither a fundamental property nor its parity since the
orientation of spin is relative to the direction of motion not to
the inertial frame as parity is. Since neutrino is the same
whether it is left-handed or right-handed if massive, there is
neither parity violation nor C violation between $\pi^{+}
\rightarrow e^{+} + \nu$ and $\pi^{-} \rightarrow e^{-} +
\overline{\nu}$ but helicity asymmetry one could call. If parity
of neutrino is violated because there is no corresponding
antiparticle, CPT is also violated in this way since it does not
hold without corresponding antiparticle.

CPT theorem states the existence of such a symmetry that satisfies
spin and statistics, but no connection between this symmetry and
its constituent discrete symmetries(C,P,T) has been verified and
the definitions of its constituent symmetries and thus CPT are
arbitrary\cite{Luders}. This theorem is proved based on
assumptions of Lorentz invariance of vacuum expectation values and
free fields and their local interactions\cite{Greenberg}. The
introduction of infinity from the assumption of free fields makes
it seemingly plausible to consider interactions as physical events
over finite time, but this requires excessive physical
interpretations that may not be necessary for instant physical
events. The interactions of fields are represented by vacuum
expectation values and they are expected to be Lorentz
invariant\cite{LSZ}. Its requirement of Lorentz invariance is
induced from field commutators of free field propagators with the
assumption of microscopic causality, but field commutators are
limited as an approximation of averaged fields over finite time
since it is inconsistent with field interpretation as a
measurement at one point for causality\cite{BohrRosenfeld}. Vacuum
expectation values are considered to be physical observables and
thus required to be Lorentz invariant. However, practical physical
observables are decay rates, which are proportional to the squared
vacuum expectation values so that they are independent of whether
vacuum expectation values are symmetric or antisymmetric. Also,
their analogies as propagators fail in interactions, since
interactions or their measurements are not expected to propagate
from one point to another but to be invariant in each spacetime
requiring all spacetimes to be the same, and fields with different
masses could make interactions as in weak interactions while
propagators require them to have the same mass. Therefore, it is
rather decay rates than vacuum expectation values that should be
physical observable and required to be Lorentz invariant. However,
we are still inconclusive on physical requirement of vacuum
expectation values; besides doubts on parity violation, the
assumption of free fields and local interactions may oversimplify
interactions; The assumption of free fields is only approximately
valid since free quarks are unavailable and composite particles
such as mesons could be different from free fields in their
interactions and underlying structures. The assumption of final
states after interaction as the unitary transformation of initial
ones \cite{LippmanSchwinger,GellMannGoldberger} is doubtful;
quarks obtained from weak interactions could different from the
initial quarks in masses and charges, photons from electron and
positron annihilation are not fermions but bosons. There must be a
set of principles for interactions even if they are instantaneous,
but the requirement of physical interpretation should be placed
with care considering the limits of fundamental assumptions while
some unphysical concepts are allowed generally in quantum field
theory: virtual photon and imaginary mass for Hermitian physics
and nonzero self interactions. Even if they are infinitesimal,
these theoretical remedies for calculations should not be
justified to be rigorous. Though the mathematical verification of
CPT theorem is doubtful, the consequences of CPT theorem could be
considered to be a fundamental assumption as pauli's principle is.

When C violation was suggested, it was shown that the lifetime of
particle and antiparticle is the same in spite of this
violation\cite{LeeYang57}. This implied that particle and
antiparticle symmetry, the mass and lifetime equality, is related
to CPT instead of C\cite{Bareboim,Murayama,McKeown}. However,
following its proof, we can even generalize this theorem to prove
that the lifetime of particle would be the same whether any
symmetry including CPT is violated or not as long as interaction
terms are the same for particle and antiparticle. Though their
proof is insufficient in the sense the application of symmetries
and definitions of particles and fiels are not well defined when
they are violated, its consequences are still valid from practical
calculations of decay rates. Therefore, it fails to establish any
connections between CPT and the mass and lifetime equality when C
is violated. It would be more appropriate to call {\it particle
and antiparticle symmetry} when one implies the mass and lifetime
equality rather than {\it CPT symmetry}, which concerns CPT
theorem for commutation relations satisfying spin and statistics.
Particle and antiparticle symmetry and CPT theorem could be
accepted as fundamental assumptions of quantum field theory
without providing a mathematical proof derived from supposedly
more fundamental assumptions. Therefore, quantum field theory
could be well established without mathematical verification of CPT
theorem.

\section{Quark Mixing}

In the standard model, quark mixing is expressed in terms of
Cabibbo-Kobayash-Maskawa (CKM) matrix, $3 \times 3$ unitary matrix
that connects the weak eigenstates and the corresponding mass
eigentates\cite{Cabibbo,Kobayashi,GIM}. However, the existence of
the weak eigenstates has not been verified and no feasible
experimental test for its verification is available. The mass
eigentstates, physical ones, are described in a matrix
representation as if it were consist of quarks, but the analogy of
rotation on quark states is insufficient for its justification
since there is no physical quantum state corresponding to it and,
in practical calculation the interactions of quarks are described
by those of individual quarks, not those of a combination of these
quarks. CKM matrix could have been investigated as coupling
constants of individual quarks first, but this has been neglected
for the unification of interactions and CP violation. The
unification of interactions suggests that weak interactions should
be described by a single coupling constant, but this is only an
interesting theory yet to be verified. The phase of CKM matrix
allowed in a matrix representation has been considered as a
possible origin of CP violation, but it is inconsistent with the
interpretation of CKM matrix elements as coupling constants since
this implies a complex coupling constant, which violates the
conservation of charge. While the unitarity of CKM matrix is
required in quark mixing, it is not if the elements of CKM matrix
are coupling constants. As the violation of this unitarity is
shown in recent experimental results\cite{Abele}, this perspective
of CKM matrix as coupling constants should be considered for its
analysis.

\section{CP Violation}
CP violation theory is based on the effective Hamiltonian from the
Weisskopf-Wigner approximation \cite{WW,LeeYangOehme}. The time
evolution of the neutral meson(denoted by $P^0$,
$\overline{P}^{0}$) can be described by
\begin{eqnarray}
i{d \over {d\tau}} \Psi
  =  (M -  {i \over 2} \Gamma)\Psi \nonumber
\end{eqnarray}
where two hermitian matrices M and $\Gamma$ in the basis of $P^0$,
$\overline{P}^{0}$ are
\begin{eqnarray}
M = \left( \begin{array}{c c} M_{11} & M_{12} \\ M^{*}_{12} &
M_{22}
\end{array} \right) ~~~\mathrm{and}~~~
\Gamma = \left( \begin{array}{c c} \Gamma_{11} & \Gamma_{12} \\
\Gamma^{*}_{12} & \Gamma_{22}
\end{array} \right) \nonumber
\end{eqnarray}
The matrix $M -  {i \over 2} \Gamma$ has eigenvalues
\begin{eqnarray}
M_{S,L} - {i \over 2} \Gamma_{S,L} &=& M_{0} - {i \over 2}
\Gamma_{0} \nonumber \\
&\pm& \sqrt{(M_{12}-{i \over 2}\Gamma_{12})(M^{*}_{12}-{i \over
2}\Gamma^{*}_{12})}  \nonumber
\end{eqnarray}
where $M_{11} = M_{22} \equiv M_{0}$ and $\Gamma_{11} =
\Gamma_{22} \equiv \Gamma_{0}$ under the assumption of particle
and antiparticle symmetry invariance. The corresponding
eigenstates are given by
\begin{eqnarray}
|P_{S}\rangle & = & [(1 + \epsilon_{P} + \delta_P)|{P^0}\rangle
+(1 - \epsilon_{P} -
\delta_{P})|{\overline{P}^{0}}\rangle]/{\sqrt2}
\quad  \nonumber\\
|{P_{L}} \rangle & = & [(1 + \epsilon_P - \delta_P)|{P^0}\rangle
-(1 - \epsilon_P + \delta_P)|{\overline{P}^{0}}\rangle]/{\sqrt 2}
\quad  \nonumber
\end{eqnarray}
The complex parameter $\epsilon_P$ represents a CP violation with
T violation, while the complex parameter $\delta_P$ represents a
CP violation with CPT violation(accurately, particle and
antiparticle symmetry). These parameters are expected to be zero
at initial time for orthogonality since symmetry violations come
from weak interactions later in time.

Though CP violation theory has been successful, there are some
problems that should be resolved in order to be consistent; The
effective Hamiltonian of neutral meson is inconsistent with
Weisskopf-Wigner approximation since initial quantum states are
presumably assumed to be eigenstates in Weisskopf-Wigner
approximation from the start while those of the effective
Hamiltonian are not. There is no proper definitions of
off-diagonal elements in the decay matrix($\Gamma$) since from
which particle they are decayed is ambiguous. CP violation phase
of CKM matrix is not necessary for CP violation and any phases
could be allowed, since the effective Hamiltonian is non-hermitian
even without CKM matrix phase. From the perspective of viewing CKM
matrix elements as individual coupling constants, this phase means
a complex coupling constant for fundamental interactions, which
makes weak interactions violate the conservation of charge. It is
also not clear how physical particles evolve in time with their
own eigenvalues while they are not orthogonal which implies one
state can be observed as the other state at any give
time\cite{DassGrimus}.

Because mesons are not point particles but composite particles, we
have some decays that could be from either a meson or its
antimeson such as $2\pi$ and $3\pi$ decays. In semileptonic
decays, only one of quark is dominating interactions in a way that
final states have the same properties of initial particles so that
they are distinguishable as in interactions of elementary particle
while, in $2\pi$ and $3\pi$ decays, both quark and antiquarks
involve so that initial mesons in decay amplitude are
indistinguishable from final states. It is these decays that are
responsible for CP violation since, if we have only semileptonic
decays, there would be no interference terms when decay rates are
summed over. Therefore, CP violation could be explained from the
fact that neutral mesons are not fundamental particles but
composite particles consist of quark and antiquark with underlying
structure.

\section{CP Violation With Effective mass}
Neutral mesons produced by strong interactions are successfully
described by approximation of two free quarks, but if mesons are
produced by weak interactions in neutral meson oscillations, its
underlying structure could make its mass effectively different
from that of meson from strong interactions. Let us consider, for
example, a neutral meson produced from strong interactions as a
ground state. Its antimeson, another quantum state, is obtained by
neutral meson oscillation exchanging quark to antiquark and vice
versa. The exchange of quarks makes its electric dipole moment
reversed. Though its total charge remains the same, this anitmeson
state could be different from the ground state and thus have
different mass since it is more likely to be unstable considering
valence quarks interacting with other ones in its underlying
structure. Though its mass is effectively different, its decays
and other interactions should be the same since they are
determined by its valence quarks. Since the mass of neutral mesons
produced by strong interactions is the same, particle and
antiparticle symmetry is valid. Here CP violation effects will be
explained by this effective mass based on Shr$\rm{\ddot{o}}$dinger
equation. Though no fundamental symmetry violations including CP
violation are implied, it will be referred to as {\it CP violation
with effective mass} for convenience. Among other possible
theoretical predictions with different choices of phase, here most
interesting calculations and their experimental tests will be
illustrated.

Let us consider two close quantum states of neutral mesons
ignoring other possible states. The Shr$\rm{\ddot{o}}$dinger
equation of the neutral meson is given by
\begin{eqnarray}
i{d \over {d\tau}} \Psi =  M\Psi \nonumber
\end{eqnarray}
where hermitian matrix M in the basis of $P^0$, $\overline{P}^{0}$
is
\begin{eqnarray}
M = \left( \begin{array}{c c} M_{0} - \Delta E/2 & M^{*}_{12}
\\ M_{12} & M_{0} + \Delta E/2
\end{array} \right) \nonumber
\end{eqnarray}
for the initial state(the ground state) is $P^0$.
\begin{eqnarray}
\overline{M} = \left( \begin{array}{c c} M_{0} + \Delta E/2 &
M_{12} \\ M^{*}_{12} & M_{0} - \Delta E/2
\end{array} \right) \nonumber
\end{eqnarray}
for the initial state(the ground state) is $\overline{P}^0$. The
matrix $M$ and $\overline{M}$ has eigenvalues
\begin{eqnarray}
M_{S,L} &=& M_{\overline{S},\overline{L}} = M_{0} \pm \sqrt{\Delta
E^{2}/4 + |M_{12}|^{2}}. \nonumber
\end{eqnarray}
In general, the corresponding eigenstates are given by
\begin{eqnarray}
|S\rangle & = & [(1 - \delta_P)|{P^0}\rangle +(1 +
\delta_{P})|{\overline{P}^{0}}\rangle]/{\sqrt {2
}}
\quad  \nonumber\\
|L \rangle & = & [(1 + \delta^{*}_P)|{P^0}\rangle -(1 -
\delta^{*}_P)|{\overline{P}^{0}}\rangle]/{\sqrt {2
}}~~~~~ \nonumber
\end{eqnarray}
for the initial state of $P^0$ and
\begin{eqnarray}
|\overline{S}\rangle & = & [(1 + \delta^{*}_P)|{P^0}\rangle +(1 -
\delta^{*}_{P})|{\overline{P}^{0}}\rangle]/{\sqrt {2
}}
\quad  \nonumber\\
|\overline{L} \rangle & = & [-(1 - \delta_P)|{P^0}\rangle +(1 +
\delta_P)|{\overline{P}^{0}}\rangle]/{\sqrt {2
}}~~~~~ \nonumber
\end{eqnarray}
for the initial state of $\overline{P}^0$. The real part of new CP
violation parameter $\delta_P$ represents the mass difference
between the ground state and the first excited state and its
imaginary part corresponds to a phase allowed in two state
Hermitian matrix.
\begin{eqnarray}
\delta_{P} = {{\Delta E/2} + i{\rm{Im}M_{12}} \over
{{\rm{Re}}M_{12} + \sqrt{ \Delta E^{2}/4 + |M_{12}|^{2}} }}
\nonumber
\end{eqnarray}
From the eigenstates we obtained, we can set up the effective
Hamiltonian
\begin{eqnarray}
i{d \over {d\tau}} \Psi
  =  (M -  {i \over 2} \Gamma)\Psi \nonumber
\end{eqnarray}
where two hermitian matrices M and $\Gamma$ in the basis of
$P^0_{S}$, ${P}^{0}_{L}$ are
\begin{eqnarray}
M = \left( \begin{array}{c c} M_{S} & 0 \\ 0 & M_{L}
\end{array} \right) ~~~\mathrm{and}~~~
\Gamma = \left( \begin{array}{c c} \Gamma_{S} & 0 \\ 0 &
\Gamma_{L}
\end{array} \right) \nonumber
\end{eqnarray}
$\Gamma_{S,L}$ will obtained from decay rates explaining the
difference in lifetimes. The initial states will evolve in time to
be
\begin{eqnarray}
|P(t) \rangle & = & (1 - \delta^{*}_{P})e^{-im_{S}t-\gamma_{S}t/2}
|S \rangle
\nonumber \\
&+&(1 + \delta_{P}) e^{-im_{L}t-\gamma_{L}t/2}| L\rangle
\nonumber \\
|\overline{P}(t) \rangle & = & (1 -
\delta_{P})e^{-im_{S}t-\gamma_{S}t/2} |\overline{S} \rangle
\nonumber \\
&+& (1 + \delta^{*}_{P}) e^{-im_{L}t-\gamma_{L}t/2}|
\overline{L}\rangle \nonumber
\end{eqnarray}
where $\Delta m = \Delta \overline{m}$, $m_{S,L} =
m_{\overline{S},\overline{L}}$ and $\gamma_{S,L} =
\gamma_{\overline{S},\overline{L}}$.

The decays of neutral mesons can be classified into two types: $2
\pi$-like decays, where its transition amplitude is contributed by
$P^{0}$ and $\overline{P}^{0}$, the others, only contributed by
either of both such as semileptonic decays. The semileptonic decay
rates for $P^0$ will be
\begin{eqnarray}
R_{f}(t) &=& {|F_{f}|^2 \over 4} \Big[(1 - 4
{\mathrm{Re}}\delta_P) e^{-\gamma_{S}t} \nonumber \\
&+& (1 + 4{\mathrm{Re}} \delta_P)e^{-\gamma_{L}t} + 2 \cos \Delta
m t e^{-\gamma t/2} \Big] \nonumber \\
R_{\overline f}(t) &=& {|F_{f}|^2 \over 4} \Big[e^{-\gamma_{S}t} +
e^{-\gamma_{L}t} - 2 \cos \Delta m t e^{-\gamma t/2} \Big]
\nonumber
\end{eqnarray}
to the first order of CP violation parameters. The first will be
referred to as the right sign decay and the later as the wrong
sign decay. The semileptonic decay rates for $\overline{P}^0$ will
be the same as before respectively for the right and wrong decays.
In $2 \pi$ decays, assuming its decay amplitudes are the same for
$P^{0}$ and $\overline{P}^{0}$, the decay rates will be
\begin{eqnarray}
R_{2\pi}(t) &=&{|F_{2\pi}|^2} \Big[(1 - 2 {\rm{Re}}
\delta_P)e^{-\gamma_{S}t} +
|\delta_P|^{2} e^{-\gamma_{L}t} \nonumber \\
&+& 2\{ {\rm{Re}}\delta_P \cos \Delta m t + {\rm{Im}}\delta_P \sin
\Delta m t\} e^{-\gamma t/2} \Big] \nonumber \\
\overline{R}_{2\pi}(t) &=&{|F_{2\pi}|^2} \Big[(1 - 2 {\rm{Re}}
\delta_P)e^{-\gamma_{S}t} +
|\delta_P|^{2} e^{-\gamma_{L}t} \nonumber \\
&+& 2\{ {\rm{Re}}\delta_P \cos \Delta m t- {\rm{Im}}\delta_P \sin
\Delta m t\} e^{-\gamma t/2} \Big] \nonumber
\end{eqnarray}
The decay rates in $3 \pi$ decays can be obtained in the same way
assuming its decay amplitudes have the opposite sign for $P^{0}$
and $\overline{P}^{0}$. Though the interference terms$({\rm{Re}}
\delta_{P} \cos \Delta m \pm {\rm{Im}} \delta_{P} \sin \Delta m)$
in $2 \pi$ are different from that of usual CP violation($\pm \cos
\Delta m$), they can be easily fit to experimental data with
proper choice of phase. Therefore, most of experimental
observations will be successfully explained by CP violation with
effective mass. The difference between the lifetimes of $P_S$ and
$P_L$ can be explained from decay rates. Let us consider only
semileptonic decays($f$ and $\overline f$) and $2 \pi$ and $3 \pi$
decays for simplicity.
\begin{eqnarray}
\Gamma_{S} &=& (|F_{2\pi}|^{2} + {|F_{f}|^{2} \over 2})(1 -
2{\rm{Re}}\delta_{P}) +|F_{3\pi}|^{2}|\delta_{P}|^{2}
 \nonumber \\
\Gamma_{L} &=& (|F_{3\pi}|^{2} + {|F_{f}|^{2} \over 2} )(1 +
2{\rm{Re}}\delta_{P})+|F_{2\pi}|^{2}|\delta_{P}|^{2} \nonumber
\end{eqnarray}
In $K^{0}$ meson, for example, $2\pi$ decays are dominating over
the other decays and $\Gamma_{S}$ is roughly $10^3$ times larger
than $\Gamma_{L}$. Since $\delta_{K}$ is about $10^{-3}$, this
predicts $2\pi$ branching ratio of $K_{L}$ to be order of
$10^{-3}$ and $3\pi$ one of $K_{S}$ to be order of $10^{-9}$.
Though experimental results for $3\pi$ in $K_{S}$(order of
$10^{-7}$) is larger than this prediction, they are in good
agreement for its simple estimation\cite{PDG}.

Though various experimental tests for this theory could be
suggested, there are normalization issues to be resolved between K
and B physics since theoretical assumptions for experimental
measurements are
conflicting\cite{JCYoonPreprint1,BaBar,Belle,CPLEARPR}. The most
distinctive test for this theory would be charge asymmetry of
$\overline{K}^{0}$, since it predicts charge asymmetry of $K^{0}$
and $\overline{K}^{0}$ to be symmetric while usual CP violation
theory does not and their difference would be order of $10^{-3}$.

\section{Conclusion}
We have discussed the limits of symmetry studies from the fact
that interacting Lagrangian fails to represent practical
calculations of interactions accurately. Parity of massive
particle should have been considered as a kind of Lorentz
invariance and thus Lorentz violations is implied by parity
violation. The approximation of taking the massless limit in the
standard model for the unification of electroweak interactions was
proved to be Lorentz violating. The interpretation of CKM matrix
as coupling constants should be considered as recent experimental
tests against the unitarity of CKM matrix are reported. Since
decays responsible for CP violation are allowed because meson is
not point particle, this could be the origin of CP violation. CP
violation is explained from composite properties of meson and
possible experimental tests is suggested. Considering the limits
and inconsistencies of theoretical assumptions and physical
observations, fundamental symmetry violation is still
inconclusive.

\end{document}